\documentclass[12pt,fleqn]{article}

\usepackage{amsmath,amssymb}
\usepackage{graphicx}
\usepackage[top=0.75in, bottom=0.75in, left=0.75in, right=0.75in]{geometry}
\usepackage[small]{caption}
\usepackage[noblocks]{authblk}
\usepackage{hyperref}
\usepackage{cite}

\usepackage{xcolor}
\usepackage{ulem}

\newcommand{\eg}{\textit{e.g.,}}

\begin{document}

\title{\sf Correlations Between the Strange Quark Condensate, Strange Quark Mass, and Kaon PCAC Relation}

\author[1]{D. Harnett\thanks{derek.harnett@ufv.ca}}
\author[2]{J. Ho\thanks{jason.ho@dordt.edu}}
\author[3]{T.G. Steele\thanks{tom.steele@usask.ca}}

\affil[1]{Department of Physics, University of the Fraser Valley, Abbotsford, BC, V2S~7M8, Canada}
\affil[2]{Department of Physics, Dordt University, Sioux Center, Iowa, 51250, USA}
\affil[3]{Department of Physics and
Engineering Physics, University of Saskatchewan, Saskatoon, SK,
S7N~5E2, Canada}

\maketitle

\begin{abstract}
Correlations between the strange quark mass, strange quark 
condensate $\langle \bar s s\rangle$, and  the kaon 
partially conserved axial current
(PCAC) relation are developed.   
The key  dimensionless and renormalization-group invariant quantities in  these correlations 
are the ratio of the strange to non-strange quark mass $r_m=m_s/m_q$, the condensate ratio  $r_c=\langle \bar s s\rangle/\langle \bar q q\rangle$,
and the kaon PCAC deviation parameter $r_p=-m_s\langle \bar s s+\bar q q\rangle/2f_K^2m_K^2$.
The correlations define a self-consistent trajectory in the $\{r_m,r_c,r_p\}$ parameter space  constraining strange quark parameters that can be used to  assess the  compatibility of different predictions of these parameters.  
Combining the constraint with Particle Data Group (PDG) values of $r_m$ results in $\{r_c,r_p\}$ constraint trajectories that are used to assess the self-consistency of various theoretical determinations of $\{r_c,r_p\}$.  The most precise determinations of $r_c$ and $r_p$ are shown to be mutually consistent with the constraint trajectories  and provide   improved bounds  on $r_p$. In general, the constraint trajectories combined with $r_c$ determinations tend to provide more accurate  bounds on $r_p$ than direct determinations. The $\{r_c,r_p\}$ correlations provide a natural identification of a self-consistent  set of strange quark mass and strange quark condensate parameters.
\end{abstract}


QCD sum-rules techniques  probe hadronic properties via QCD composite operators with  appropriate quantum numbers and quark/gluonic valence content to serve as interpolating fields for hadronic states~\cite{Shifman:1978bx,Shifman:1978by} (for reviews, see \eg\ Refs.~\cite{Reinders:1984sr,Narison:2007spa,Gubler:2018ctz}).  Correlation functions of these composite operators are calculated from QCD using the operator-product expansion and are then related  to hadronic properties through dispersion relations, which are converted to a QCD sum-rule through application of an appropriate transform.    Important features of these correlation functions are the power-law corrections induced by QCD condensates (vacuum expectations values) that parameterize non-perturbative aspects of the QCD vacuum.  

In QCD sum-rules analyses of hadronic systems containing strange quarks, 
the strange quark mass $m_s$ and the strange quark condensate  $\langle \bar s s\rangle$ are important parameters. 
Depending upon the system, the condensate may emerge as  $\langle \bar s s\rangle$ or be 
accompanied with quark mass factors (\eg\  $m_s\langle \bar s s\rangle$, $m_c\langle \bar s s\rangle$).  However,  
$\langle \bar s s\rangle$ is also used within determinations of higher-dimension condensates, including the vacuum-saturation approximation for dimension-six quark condensates and the dimension-five mixed condensate  \cite{Beneke:1992ba,Belyaev:1982sa}
\begin{equation}
\frac{1}{2} \langle g\bar s\sigma_{\mu\nu} \lambda^a G^a_{\mu\nu}  s\rangle=\langle g\bar{s} \sigma G s\rangle=
\left((0.8\pm 0.1)\ {\rm GeV^2}\right) \langle \bar s s\rangle ~.
\label{mixed_condensate}
\end{equation}
Because of these multiple roles in determining different QCD condensates,  the strange quark condensate is an essential parameter in QCD sum-rules and provides insight into $SU(3)$ flavour symmetry of the QCD vacuum.

QCD sum-rules studies of the strange quark condensate are based upon pseudoscalar and scalar correlation functions combined with low-energy theorems (see \eg\ Refs.~\cite{Narison:1981mz,Dominguez:1985vc,Dominguez:2018zzi}) 
\begin{gather}
\Psi_5\left(q^2\right)=i\int d^4x \left\langle \Omega\right\vert T(J_5(x) J_5^\dagger(0) \left\vert \Omega\right\rangle\,,
~
J_5(x)=i\left(m_u+m_s\right)\bar s(x) \gamma_5 u(x)\,
\\
\Psi_5(0)=-\left(m_s+m_u\right)\langle \bar s s+\bar q q \rangle\,
\label{LET_PS}
\\
\Psi\left(q^2\right)=i\int d^4x \left\langle \Omega\right\vert T(J(x) J^\dagger(0) \left\vert \Omega\right\rangle\,,
~J(x)=i\left(m_s-m_u\right)\bar s(x) u(x)\,
\\
\Psi(0)=-\left(m_s-m_u\right)\langle \bar s s-\bar q q \rangle\,
\label{LET_S}
\end{gather} 
where $SU(2)$ isospin symmetry of the vacuum implies
$\langle \bar q q \rangle=\langle \bar u u\rangle=\langle \bar d d \rangle$.

QCD sum-rules determinations  of the low-energy constants $\Psi_5(0)$ and $\Psi(0)$ contain  $\langle \bar s s\rangle$ dependence and thus  provide a natural means to extract the strange quark condensate (see \eg\ Refs.~\cite{Narison:1981mz,Dominguez:1985vc}). For example, the ratio $\Psi_5(0)/\Psi(0)$ can be used to reference 
the strange quark condensate to the non-strange condensate through the renormalization-group (RG) invariant ratio and the RG-invariant strange quark mass ratio $\xi=m_u/m_s=0.024\pm 0.006\ll 1$ \cite{Tanabashi:2018oca}
\begin{gather}
\frac{\Psi_5(0)}{\Psi(0)}=\left(\frac{1+\xi}{1-\xi}\right)\left(\frac{r_c+1}{r_c-1}\right)\approx \frac{r_c+1}{r_c-1}
\label{LET_ratio}
\\
\frac{\langle \bar s s \rangle}{\langle \bar  qq \rangle}=r_c\,,~\langle \bar q q \rangle=\langle \bar u u\rangle=\langle \bar d d \rangle\,.
\label{r_c}
\end{gather}
Furthermore, as $\langle\bar{q}q\rangle<0$, 
if $r_c<1$ then $\Psi(0)<0$, so the 
sign of the low-energy constant $\Psi(0)$ provides valuable information on qualitative aspects of $SU(3)$ flavour 
symmetry breaking by the vacuum.

\begin{figure}[htb]
\centering
\includegraphics[scale=1]{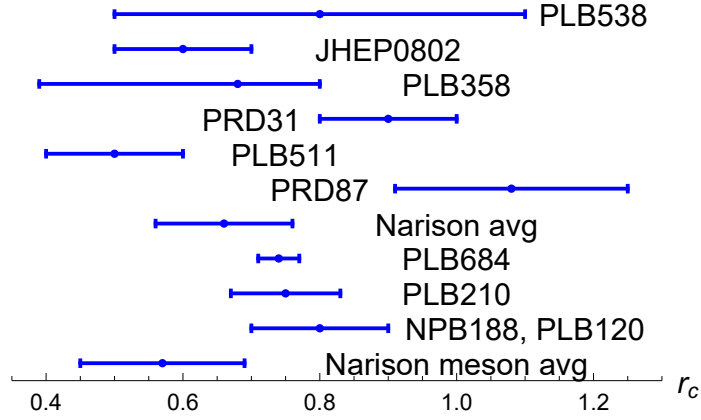}
\caption{Summary of $r_c$ theoretical determinations.  See Table \ref{rc_rp_table} for key to references.
}
\label{rc_fig}
\end{figure}

There  is a wide range of $r_c$ theoretical determinations.  
QCD sum-rules analyses of light  scalar and pseudoscalar meson correlation
functions have been used to determine 
$r_c$~\cite{Dominguez:1985vc,Narison:1995hz,Dominguez:2001ek,Dominguez:2007hc},
and a combined estimate from these results is $r_c=0.57\pm 0.12$ \cite{Narison:2007spa}. 
QCD sum-rules for baryon mass splittings tend to yield larger values ranging from the earliest determinations  
$r_c=0.8\pm 0.1$~\cite{Ioffe:1981kw,Reinders:1982qg} to the updated values 
$r_c=0.75\pm 0.08$~\cite{Narison:1988ep} 
and  $r_c=0.74\pm 0.03$~\cite{Albuquerque:2009pr}. 
Because the QCD sum-rules for heavy baryon mass splittings have better perturbative convergence compared to the light meson analyses, the most precise sum-rule value is
$r_c=0.74\pm 0.03$~\cite{Albuquerque:2009pr}.
 A more conservative sum-rule value  obtained from an average of light meson and baryon systems 
is $r_c=0.66\pm 0.10$~\cite{Narison:2007spa}.
The lattice QCD determination $r_c=1.08\pm 0.16$~\cite{McNeile:2012xh} 
(multiple sources of uncertainty have been combined) is
considerably larger than the sum-rules determinations. 
An analysis combining aspects of QCD sum-rules and lattice QCD results yields $r_c=0.8\pm 0.3$~\cite{Jamin:2002ev}. 
The various theoretical determinations of $r_c$ are shown in Figure~\ref{rc_fig} and Table~\ref{rc_rp_table}.

\begin{table}[htb]
\centering
\begin{tabular}{||c|c|c|c||}
\hline
$r_p$ & $r_c$ & Reference &  Methodology\\
\hline\hline
$0.56\pm 0.16$ & $0.5\pm 0.1$ & \cite{Dominguez:2001ek} (PLB511)  & LSR (four loops)
\\
\hline
$0.66^{+0.23}_{-0.17}$ & $0.68^{+0.15}_{-0.29}$ & \cite{Narison:1995hz} (PLB358) & QCDSR
\\
\hline
$0.57\pm 0.19$ & $0.66\pm 0.1$ & \cite{Narison:2007spa} (Narison avg) & QCDSR average 
\\
\hline
$0.66\pm 0.05$ & $0.6\pm 0.1$ & \cite{Dominguez:2007hc} (JHEP0802) & FESR five loop
\\
\hline
$0.39\pm 0.22$ & $0.8\pm 0.3$ & \cite{Jamin:2002ev} (PLB538)  & ChPT, LSR, and LQCD input
\\
\hline
$0.74\pm 0.16$ & $1.08\pm 0.16$ & \cite{McNeile:2012xh} (PRD87)  & LQCD 
\\
\hline
--- & $0.9\pm 0.1$ & \cite{Dominguez:1985vc} (PRD31)  & LSR
\\
\hline
--- & $0.74\pm 0.03$ & \cite{Albuquerque:2009pr} (PLB684)  & LSR heavy baryons 
\\
\hline
--- & $0.75\pm 0.08$ & \cite{Narison:1988ep} (PLB210)  & QCDSR heavy mesons
\\
\hline
--- & $0.8\pm 0.1$ & \cite{Ioffe:1981kw,Reinders:1982qg} (NPB188, PLB120)  & QCDSR baryons
\\
\hline
--- & $0.57\pm 0.12$ & \cite{Narison:2007spa} (Narison meson avg)  & QCDSR meson average
\\
\hline
$0.5\pm 0.17$ & --- & \cite{Dominguez:1984yx} (ZPC27)  & LSR, AQCD
\\
\hline
$0.63\pm 0.08$ & --- & \cite{Narison:1981mz} (PLB104)  & QCDSR
\\
\hline
\hline
\end{tabular}
\caption{Summary of \{$r_c$, $r_p$\} 
theoretical determinations shown in Figs.~\ref{rc_fig}, \ref{rp_fig}  and \ref{corr_error_fig}. In some entries, multiple sources of uncertainty have been combined or ranges have been converted to an uncertainty. FESR denotes Finite-Energy sum-rules, LSR denotes Laplace sum-rules, ChPT denotes Chiral Perturbation theory, QCDSR denotes QCD sum-rules, AQCD denotes analytic continuation by duality, and LQCD denotes Lattice QCD. The PLB511 entry includes a truncation uncertainty from  Ref.~\cite{Narison:2007spa}.}

\label{rc_rp_table}
\end{table}
The PCAC Gell-Mann-Oakes-Renner (GMOR) relation relates   
the RG-invariant combination of the non-strange quark condensate and mass parameter to  pion properties 
\cite{GellMann:1968rz}
\begin{equation}
2m_q\langle \bar qq\rangle =-f_\pi^2m_\pi^2\,,~m_q=\frac{m_u+m_d}{2}~,
\label{GMOR}
\end{equation} 
where in our conventions $f_\pi=130/\sqrt{2}\,{\rm MeV}$ \cite{Tanabashi:2018oca}.
Deviations from the pion PCAC relation are bounded by approximately 5\% \cite{Jamin:2002ev,McNeile:2012xh,Dominguez:2018zzi} and are thus a small numerical effect.
The RG-invariant  quark mass ratio \cite{Tanabashi:2018oca}
\begin{equation}
r_m=\frac{m_s}{m_q} = 27.3\pm 0.7
\label{rm_pdg}
\end{equation}
can then be combined with $r_c$ to obtain the following strange quark condensate $m_s\langle\bar s s\rangle$ 
in terms of~\eqref{GMOR}
\begin{equation}
m_s\langle\bar s s\rangle =r_m r_c m_q \langle \bar qq\rangle\,.
\label{mss_ratio}
\end{equation}

\begin{figure}[htb]
\centering
\includegraphics[scale=0.9]{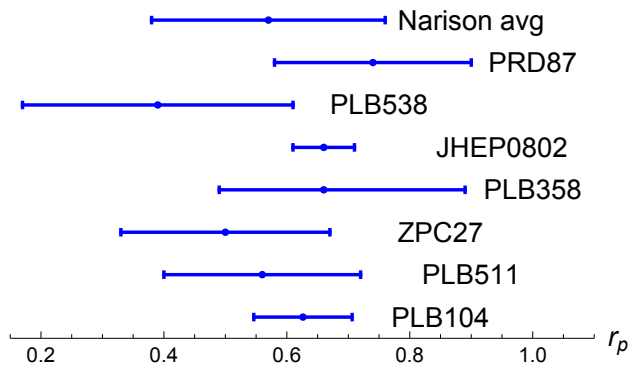}
\caption{Summary of $r_p$ theoretical determinations. See Table \ref{rc_rp_table} for key to references.
}
\label{rp_fig}
\end{figure}

A complementary  approach to determinations of $m_s\langle \bar s s\rangle$ is through the deviation from the kaon PCAC relation
as parameterized by $r_p$
\begin{equation}
-\left(m_s+m_u\right)\langle \bar q q +\bar s s\rangle \approx -m_s\langle \bar q q +\bar s s\rangle =r_p 2 f_K^2m_K^2\,,
\label{pcac_deviation}
\end{equation} 
where  $r_p=1$ corresponds to the kaon PCAC result (in  conventions where $f_K=156/\sqrt{2}\,{\rm MeV}$ \cite{Tanabashi:2018oca}) and \eqref{rm_pdg} implies that neglecting $m_u$  is a small numerical effect 
(see \eg\ Eq.~\eqref{LET_ratio}).  
The PCAC deviation parameter $r_p$ can be determined in QCD sum-rules by using the low-energy theorem for the pseudoscalar correlation function \eqref{LET_PS} (see \eg\ Ref.~\cite{Narison:1981mz}).
A sum-rule evaluation of $\Psi_5(0)$ thus allows determination of $r_p$  via Eqs.~\eqref{pcac_deviation} and \eqref{LET_PS}.  Significant deviations from the kaon PCAC result have been found in this approach ranging from the earliest values $r_p=0.63\pm 0.08$
\cite{Narison:1981mz} and  $r_p=0.5\pm 0.17$ \cite{Dominguez:1984yx}, 
to later determinations from Laplace sum-rules $r_p=0.57\pm 0.19$ \cite{Dominguez:2001ek}\footnote{The result of \cite{Dominguez:2001ek} has been augmented with an estimated  truncation error \cite{Narison:2007spa}. }, 
 QCD sum-rules $r_p=0.66^{+0.23}_{-0.17}$ \cite{Narison:1995hz}, 
 lattice QCD $r_p=0.74\pm 0.16$ \cite{McNeile:2012xh},
 and combined approaches (merging Laplace sum-rules, chiral perturbation theory, and lattice QCD input) $r_p=0.39\pm 0.22$ \cite{Jamin:2002ev}. The most precise determination  $r_p=0.66\pm 0.05$ emerges from finite-energy sum-rules \cite{Dominguez:2007hc}. 
The various determinations of $r_p$ are shown in Figure \ref{rp_fig}. 

\begin{figure}[htb]
\centering
\includegraphics[scale=0.9]{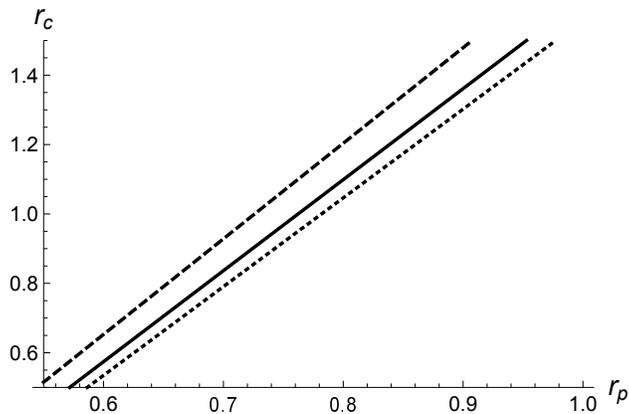}
\caption{Correlation \eqref{constraint} between $r_p$ and $r_c$ for PDG values \cite{Tanabashi:2018oca} of $r_m$ 
(solid black, dashed black, and dotted black lines respectively representing central, upper, and lower PDG values).
}
\label{corr_fig}
\end{figure}

The RG-invariant combination of the strange quark mass and condensate emerging from the kaon PCAC relation \eqref{pcac_deviation}  is
\begin{equation}
m_s\langle \bar s s\rangle= -r_p 2 f_K^2m_K^2-r_m m_q\langle \bar q q\rangle\,.
\label{mss_rp}
\end{equation} 
The two expressions \eqref{mss_ratio} and \eqref{mss_rp} for $m_s\langle \bar s s\rangle$  are thus self-consistent if the following constraint is satisfied:
\begin{gather}
r_c=\frac{r_p}{\sigma_m}-1\,,
\label{constraint}
\\
\sigma_m=r_m\left(\frac{f_\pi^2m_\pi^2}{4f_K^2m_K^2}\right)\,,
\end{gather}
providing a correlation in the $\{r_m,r_c,r_p\}$ parameter space.  The  $\{r_c,r_p\}$ linear trajectories resulting from the PDG $r_m$ range \eqref{rm_pdg} are shown in Fig.~\ref{corr_fig}.  Determinations of $r_c$ and $r_p$ that lie along these trajectories will thus be self-consistent for the PDG range of the strange quark mass. 
The correlation trajectories provide a relatively more stringent constraint on $r_p$ compared to $r_c$. For example, the conservative range $0.6<r_c<1.2$ leads to a relatively narrow range  $0.57<r_p<0.86$ corresponding to a significant deviation from the kaon PCAC relation.  

\begin{figure}[htb]
\centering
\includegraphics[scale=1]{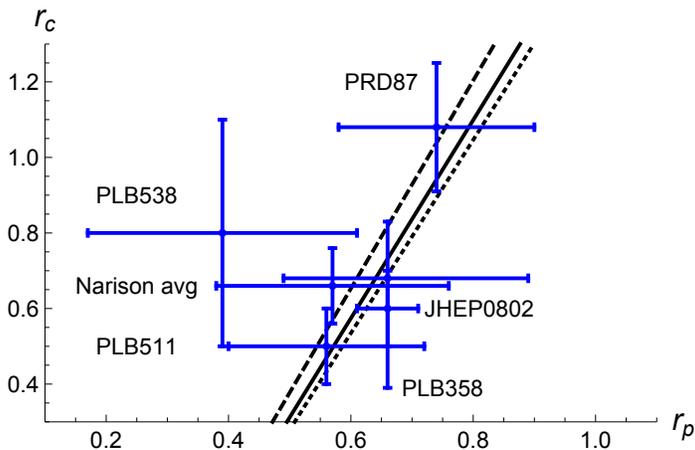}
\caption{Comparison of simultaneous \{$r_c$, $r_p$\} determinations with the \{$r_p$, $r_c$\} correlation  trajectories \eqref{constraint} respectively representing central, upper, and lower PDG $r_m$ values (solid black, dashed black, and dotted black lines). See Table \ref{rc_rp_table} for key to references.}
\label{corr_error_fig}
\end{figure}

The most interesting analyses from the literature are those which simultaneously  allow determination of both $r_c$ and $r_p$ because they map out a region in the $\{r_c,r_p\}$ parameter space that can be compared with the constraint trajectories.   In Fig.~\ref{corr_error_fig}, the simultaneous determinations of $\{r_c,r_p\}$ are compared with the linear constraint trajectories.  The  determinations from Refs.~\cite{Narison:2007spa,Dominguez:2001ek,Dominguez:2007hc,Narison:1995hz,McNeile:2012xh} 
show good agreement with the trajectories, but Ref.~\cite{Jamin:2002ev}, which has the smallest determination of $r_p$,
 does not intersect the trajectories.  
However, Ref.~\cite{Jamin:2002ev} is somewhat different than the other simultaneous determinations in Fig.~\ref{corr_error_fig} because it combines  different methodologies (the $r_p$ determination in Ref.~\cite{Jamin:2002ev} involves chiral perturbation theory 
whereas $r_c$ involves QCD sum-rules).  

\begin{figure}[htb]
\centering
\includegraphics[scale=1]{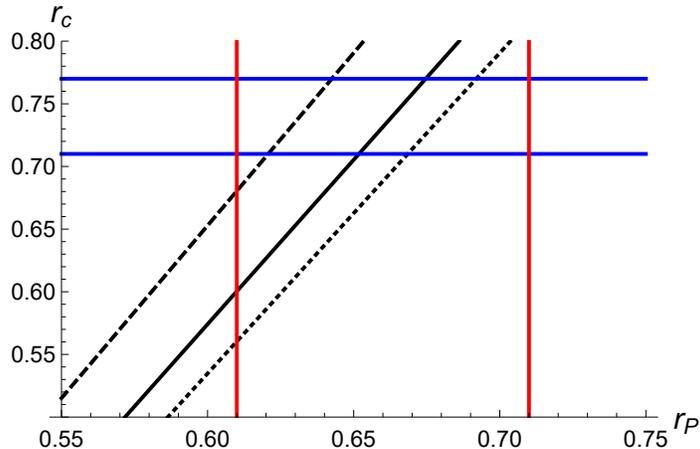}
\caption{Comparison of the Ref.~\cite{Albuquerque:2009pr} $r_c$ bounds (blue lines) and  Ref.~\cite{Dominguez:2007hc} $r_p$ bounds (red vertical lines) with the $r_p$, $r_c$ correlation  trajectories \eqref{constraint} respectively representing central, upper, and lower PDG $r_m$ values (solid black, dashed black, and dotted black lines). }
\label{rp_rc_bounds_fig}
\end{figure}

\begin{figure}[htb]
\centering
\includegraphics[scale=1.0]{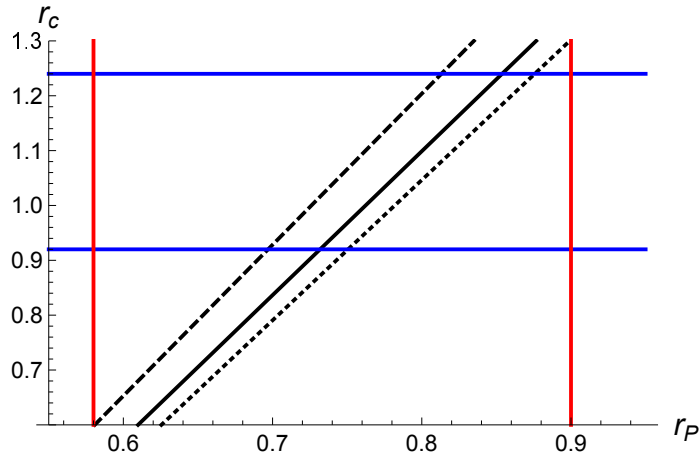}
\caption{Comparison of the Ref.~\cite{McNeile:2012xh} $r_c$ bounds (blue lines) and  $r_p$ bounds (red vertical lines) with the $r_p$, $r_c$ correlation  trajectories \eqref{constraint} respectively representing central, upper, and lower PDG $r_m$ values (solid black, dashed black, and dotted black lines).}
\label{rp_rc_bounds_lattice_fig}
\end{figure}

As a final consideration, Fig.~\ref{rp_rc_bounds_fig} assesses the most precise individual determinations  $r_c=0.74\pm 0.03$ 
\cite{Albuquerque:2009pr} and $r_p=0.66\pm 0.05$ \cite{Dominguez:2007hc} against the constraint trajectories.   The two determinations delineate a  compatible region of $\{r_c, r_p\}$ parameter space, and as discussed above, the $r_c$ determination provides a tighter bound 
$0.62<r_p<0.69$ completely contained within the Ref.~\cite{Dominguez:2007hc} determination. A similar feature for the Ref.~\cite{McNeile:2012xh} lattice determinations is shown in Fig.~\ref{rp_rc_bounds_lattice_fig}; the constraint trajectories combined with the $r_c$ determination again provides a tighter bound $0.69<r_p<0.88$ completely contained within the Ref.~\cite{McNeile:2012xh} determination.  Thus the constraint trajectories combined with $r_c$ determinations  tend to provide more accurate  bounds on $r_p$ than direct determinations.   

Based on a small set of input parameters, 
we can generate a collection of self-consistent numerical results for some QCD condensates 
containing strange quarks. 
The strange quark condensate $\langle\bar{s}s\rangle$ 
is often multiplied by a quark mass, \textit{i.e.,} $M\langle\bar{s}s\rangle$,
an RG-invariant quantity. 
Using $r_c=0.74 \pm 0.03$ and~(\ref{GMOR})--(\ref{mss_ratio}), we find
\begin{equation}\label{mss}
  m_s \langle\bar{s}s\rangle = (-1.7\times 10^{-3})\ \text{GeV}^4.
\end{equation}
Then, using
\begin{gather}
   m_c\langle\bar{s}s\rangle=
  \left(\frac{m_c}{m_s}\right)m_s\langle\bar{s}s\rangle\\
   m_b\langle\bar{s}s\rangle
   =\left(\frac{m_b}{m_s}\right) m_s\langle\bar{s}s\rangle
\end{gather}
with quark mass ratios~\cite{Tanabashi:2018oca}
\begin{gather}
  \frac{m_c}{m_s}=11.72 \pm 0.25\label{mcms_ratio}\\
  \frac{m_b}{m_s}=53.94 \pm 0.12\label{mbms_ratio},
\end{gather}
we find 
\begin{gather}
  m_c \langle\bar{s}s\rangle = (-1.9\times 10^{-2})\ \text{GeV}^4\label{mcss}\\
  m_b \langle\bar{s}s\rangle = (-9.0\times 10^{-2})\ \text{GeV}^4\label{mbss}.
\end{gather}
Uncertainties in~(\ref{mss}), (\ref{mcss}), and~(\ref{mbss}) are at most~5\% 
and stem mainly from deviations from pion PCAC in \eqref{GMOR}.
Like $\langle\bar{s}s\rangle$, 
the dimension-five mixed strange quark condensate~(\ref{mixed_condensate}) 
is often multiplied by a 
quark mass, \textit{i.e.,} $M\langle g\bar{s}\sigma G s\rangle$.
Equations~(\ref{mixed_condensate}) and~(\ref{mss}) give
\begin{equation}\label{ms_mixed}
     m_s \langle g \bar{s}\sigma Gs\rangle = (-1.3\times 10^{-3})\ \text{GeV}^6.
\end{equation}
Then, again using the quark mass ratios~(\ref{mcms_ratio}) and~(\ref{mbms_ratio}), we find 
\begin{gather}
  m_c \langle g \bar{s}\sigma Gs\rangle = (-1.6\times 10^{-2})\ \text{GeV}^6\label{mc_mixed}\\
  m_b \langle g \bar{s}\sigma Gs\rangle = (-7.2\times 10^{-2})\ \text{GeV}^6.\label{mb_mixed}
\end{gather}
Uncertainties in~(\ref{ms_mixed})--(\ref{mb_mixed}) are roughly
18\% and stem from deviations from pion PCAC~\eqref{GMOR} and from the ratio of the strange quark condensate
to the dimension-five mixed strange quark condensate in~(\ref{mixed_condensate}).
Dimension-six quark condensates are often multiplied by a factor of $\alpha_s$. 
In addition, the vacuum saturation hypothesis
is generally used to express dimension-six quark condensates as products of 
dimension-three quark condensates~\cite{Shifman:1978bx}.
Using
\begin{equation}
    \alpha_s \langle\bar{q}q\rangle^2 = 1.8\times 10^{-4}\ \text{GeV}^6\label{aqqqq}
\end{equation}
from Ref.~\cite{Shifman:1978by}, we find with $r_c=0.74$ that
\begin{gather}
  \alpha_s \langle\bar{s}s\rangle\langle\bar{q}q\rangle = 1.3\times 10^{-4}\ \text{GeV}^6\\
  \alpha_s \langle\bar{s}s\rangle^2 = 9.9\times 10^{-5}\ \text{GeV}^6\label{assss}.
\end{gather}
Deviations from vacuum saturation  
can be applied by multiplying the results~(\ref{aqqqq})--(\ref{assss}) by a 
vacuum saturation parameter $\kappa$ where $1\leq\kappa\lesssim 4$ 
(\textit{e.g.,}~\cite{Bertlmann:1987ty,Narison:1995jr}).
Note that the result~\eqref{aqqqq} is remarkably consistent with~\eqref{GMOR} 
and the PDG~\cite{Tanabashi:2018oca} 
values $m_q(2\ {\rm GeV})=3.45\ {\rm MeV}$ and $\alpha_s(2\,{\rm GeV})$.

In summary, we have developed the constraint \eqref{constraint} relating the strange quark parameters $r_m=m_s/m_q$, $r_c=\langle \bar s s\rangle/\langle \bar q q\rangle$ and the kaon PCAC deviation parameter $r_p=-m_s\langle \bar s s+\bar q q\rangle/2f_K^2m_K^2$. 
Using $r_m$ PDG \cite{Tanabashi:2018oca} values, $\{r_c,r_p\}$ theoretical determinations 
(see Table~\ref{rc_rp_table}) are compared with the constraint trajectories (see Fig.~\ref{corr_error_fig}).   Theoretical predictions corresponding to the smallest value of $r_p$ show poor agreement with the constraint trajectories.   However, Fig.~\ref{rp_rc_bounds_fig} demonstrates that the most precise 
values  $r_c=0.74\pm 0.03$ 
\cite{Albuquerque:2009pr} and $r_p=0.66\pm 0.05$ \cite{Dominguez:2007hc} are mutually consistent with the $\{r_c,r_p\}$ constraint trajectories and provide an improved determination  $0.62<r_p<0.69$.  
The combination of $\{r_c,r_p\}$ constraint trajectories with $r_c$ determinations to obtain improved $r_p$ 
bounds is also observed in 
Fig.~\ref{rp_rc_bounds_lattice_fig}  for the lattice values \cite{McNeile:2012xh}.  Thus the $\{r_c,r_p\}$ constraint trajectories provide  a valuable methodology for assessing self-consistency or improving  accuracy of  determinations of the condensate ratio  $r_c=\langle \bar s s\rangle/\langle \bar q q\rangle$
and the kaon PCAC deviation parameter $r_p=-m_s\langle \bar s s+\bar q q\rangle/2f_K^2m_K^2$.

\section*{Acknowledgments}
The work is supported by the Natural Sciences and Engineering Research Council of Canada (NSERC). 


\end{document}